\documentclass[aps,pre,preprint]{revtex4}
\usepackage{epsfig}
\begin{document}
\title{Collapse of solitary waves near transition from supercritical to subcritical bifurcations}

\author{D.S. Agafontsev $^{(a)}$, F. Dias $^{(b)}$ and E.A. Kuznetsov $^{(c,a)}$}

\affiliation{ {\small \it $^{(a)}$ L.D. Landau Institute for Theoretical Physics, 2 Kosygin str., 119334 Moscow, Russia}\\
{\small \it $^{(b)}$ Centre de Math\'ematiques et de Leurs Applications, Ecole Normale Sup\'erieure de Cachan, 61 avenue du Pr\'esident Wilson, 94235 Cachan cedex, France}\\
{\small \it $^{(c)}$ P.N. Lebedev Physical Institute, 53 Leninsky ave., 119991 Moscow, Russia}
}

\begin{abstract}
We study both analytically and numerically the nonlinear stage of the instability of one-dimensional solitons in a small vicinity of the transition 
point from supercritical to subcritical bifurcations  in the framework of the generalized nonlinear Schr\"{o}dinger equation.  It is shown that near the collapsing time the pulse amplitude and its width demonstrate the self-similar behavior with a small asymmetry at the pulse tails due to  self-steepening.  This theory is applied to both solitary interfacial deep-water waves and  envelope water waves with a finite depth and short optical pulses in fibers as well.
\end{abstract}

\maketitle

{\bf 1.}
According to the usual definition, solitons   are nonlinear localized objects propagating uniformly with a constant velocity (see, for example, \cite{ZMNP}). The soliton velocity $V$ represents the main soliton characteristics which  defines very often the soliton shape, in particular, its amplitude and width. The  velocity $V$ takes values in the bounded interval with the
end points  given by the conditions of touching
in the $k$-space of the plane $\omega=({\bf k\cdot V})$  with the dispersion relation of linear waves $\omega=\omega({\bf k})$; 
intersection of these surfaces means presence of the Cherenkov resonance that prevents the existence of stationary localized entities.
At these boundaries the soliton velocity reaches its extremal value $V_{cr}$ equal to the maximum (or minimum) phase velocity of linear waves and, as a result,  solitons there undergo a bifurcation, either supercritical or subcritical. While approaching the supercritical bifurcation point from below or above  the soliton amplitude vanishes smoothly  according to the same  - Landau - law ($\propto |V-V_{cr}|^{1/2}$) as for phase transitions of the second kind. 
The behavior of solitons in this case is completely universal, both for their amplitudes and their shapes. As $V\to V_{cr}$  solitons transform into oscillating wave trains with  the carrying frequency corresponding to the extremal phase velocity of linear waves $V_{cr}$. The shape of the wave train envelope coincides with that for the soliton of the standard - cubic - nonlinear Schr\"{o}dinger equation.  The soliton width happens to be  proportional to $|V-V_{cr}|^{-1/2}$. 

In the case of subcritical bifurcation, the situation is similar to phase transitions of the first kind: at the critical velocity the soliton undergoes a jump in its amplitude.  In this case the corresponding  theory can be developed near the transition point between  subcritical  and   supercritical bifurcations (in analogy with the tri-critical point for phase transitions).  In the series of papers \cite{DI96, K99, AFK1, AFK2} we demonstrated that in this case the soliton behavior can be described by means of the generalized nonlinear Schr\"{o}dinger equation (NLSE) for the envelope $\psi$, which in the one-dimensional case reads as follows:
\begin{eqnarray}\label{GNLSE}
i\frac{\partial \psi}{\partial t}-\lambda ^2\omega _0\psi +\frac{\omega'' _0}{2}\psi _{xx}-\mu |\psi
|^2\psi +4i\beta |\psi |^2\psi _x  +\gamma \psi \widehat k|\psi |^2+3C|\psi|^4\psi =0 \nonumber  
\end{eqnarray}
where $\omega_0\equiv \omega(k_0)$ and $k_0$ are the carrying frequency and wave number, respectively,  $\lambda ^2=(V_{cr}-V)/V_{cr}\ll 1$, $\omega'' _0$ the second derivative of $\omega(k)$ taken at $k=k_0$. Here the four-wave coupling coefficient $\mu$ is assumed to have additional smallness characterizing the proximity to the transition from supercritical to subcritical bifurcations.  The transition point is defined from the equation $\mu=0$. For example, for interfacial deep-water waves propagating along the interface between two ideal fluids in the presence of capillarity \cite{AFK1, AFK2}
\[
\mu=\frac{k_0^3}{1+\rho }\left( A_{cr}^2-A^2\right), 
\]
where $\rho$ is the density ratio, $A=(1-\rho)/(1+\rho)$ the Atwood number,  $A_{cr}^2=5/16$ and
$\rho _{cr}=(21-8\sqrt{5})/11$, as it was shown in Ref. \cite{DI96}. 
For $\rho <\rho _{cr}$, the
four-wave coupling coefficient $\mu$ is negative, and the corresponding
nonlinearity is of the focusing type. In this case, solitary waves near the
critical velocity $V_{cr}$ are described by the stationary ($\partial /\partial t=0$) NLSE 
and undergo a supercritical bifurcation at $V=V_{cr}$ \cite{DI96}. For $\rho
>\rho _{cr}$ the coupling coefficient changes sign and, as a result, the bifurcation
becomes subcritical. For water waves in finite depth $h$ the coefficient $\mu$ changes its sign at $\theta_{cr}=k_{0}h\approx 1.363$ \cite{whitham}
while $\omega''_0$ is always negative. Thus the nonlinearity belongs to the focusing type for $\theta (=kh) >\theta_{cr}$ and respectively becomes defocusing in the region $\theta<\theta_{cr}$ \cite{whitham, A3}. In nonlinear optics, as shown in \cite{K99}, a decrease of $\mu$ (``Kerr'' constant) can be provided by the interaction of light pulses with acoustic waves (Mandelstamm-Brillouin scattering). 

Because of the smallness of $\mu$  we keep in Eq. (\ref{GNLSE}) the next order nonlinear terms: the gradient term 
($\sim \beta$)  responsible for self-steepening of the pulse (analog of the Lifshitz term in phase transitions), the  nonlocal term (due to the presence of the integral operator $\widehat k$,  the Fourier transform of its kernel is equal to $|k|$) and the six-wave nonlinear term with coupling coefficient $C$. 
Two additional 4-wave interaction terms, both local and nonlocal, appear as the result of the expansion of the four-wave matrix element $T_{k_1k_2k_3k_4}$ in powers of the small parameters $\kappa_i=k_i-k_0$:
\begin{eqnarray}  \label{exprT}
&& T_{k_1k_2k_3k_4} =\frac{\mu}{2\pi} +\frac{\beta}{2\pi} (\kappa
_1+\kappa _2+\kappa_3+\kappa _4) \\
&& -\frac{\gamma}{8\pi} (|\kappa _1-\kappa _3|+|\kappa _2-\kappa
_3|+|\kappa _2-\kappa _4|+|\kappa _1-\kappa _4|).  \nonumber
\end{eqnarray} 
The existence of the nonlocal contribution in the expansion is connected with non-analytical dependence of the matrix element $T$ in its arguments. For interfacial deep-water waves (IW) this non-analyticity originates from the solution of Laplace equation for the hydrodynamic potential and its reduction to the moving interface. 
For instance, for water waves (WW) with a finite depth the nonlocal term is absent
\cite{A3} as well as for electromagnetic waves in nonlinear dielectrics \cite{K99} because of the analyticity of matrix elements with respect to frequencies, which is a consequence of causality (see, for example, Refs. \cite{LandauLifshitz, K99}).  In the latter case the spatial dispersion effects are relativistically small and can be neglected. 

For both IW and WW near the transition point,  $\omega_0''C$ is positive; moreover $\gamma$ is also positive for IW,  and therefore the  corresponding nonlinearities are focusing, thus providing the existence of localized solutions. Depending on the sign of $\mu$ there exist two branches of solitons. For IW they were found numerically \cite{AFK1, AFK2} using the Petviashvili scheme \cite{petviashvili}. Explicit solutions for both kinds of IW solitons can be obtained in the limiting case only when $V\to V_{cr}$. For negative $\mu$ these are the classical NLS solitons with a sech shape. For the subcritical bifurcation at $V = V_{cr}$ the soliton amplitude  remains finite with algebraic decay ($\sim 1/|x|)$ at infinity  \cite{AFK1, AFK2}. When nonlocal nonlinearity is absent  ($\gamma=0$) soliton solutions can be found explicitly. For both branches at large $\lambda$ the number of waves $N=\int|\psi|^2 dx$ approaches from below and above the same value 
$N_{cr}$ which coincides with the number of waves $N$ on the solitons with $\mu=0$. This property for solitons in fibers  means that
the energy of optical pulse saturates,  tending to the constant value with a decrease of the pulse duration.   

On the other hand, all solitons of Eq. (\ref{GNLSE}) are stationary points of the Hamiltonian $H$ for fixed number of waves : $\delta (H+\lambda^2 N)=0$, where the Hamiltonian in dimensionless variables is given by
\begin{eqnarray}
H=\int\bigg[|\psi_{x}|^{2}+\frac{\mu}{2}|\psi|^{4}+i\beta(\psi_{x}^{*}\psi-\psi_{x}\psi^{*})|\psi|^{2}-\frac{\gamma}{2}|\psi|^{2}\hat{k}|\psi|^{2}-C|\psi|^{6}\bigg]\,dx.  \label{hamiltonian}
\end{eqnarray}
This allows one to use the Lyapunov theorem in the analysis of their stability.
Here, for the IW, $\lambda\sim (V_{cr}-V)^{1/2}{|\rho-\rho_{cr}}|^{-1}$, $C={319}/{1281}$ and
\begin{equation} \label{IW}
\mu={\rm sign}(\rho -\rho_{cr}), \quad \beta={6}/{\sqrt{427}}, \quad \gamma={32}/{\sqrt{427}}, 
\end{equation}
and for the WW case (compare  with Ref. \cite{A3})
\begin{equation} \label{WW}
\mu={\rm sign}(\theta_{cr}-\theta),\,\,\, \beta\approx -0.397,  \,\,\, C\approx 0.176.
\end{equation}
As shown in \cite{K99, AFK1, AFK2, A3}, for $N<N_{cr}$ solitons corresponding to the supercritical branch realize the minimum values of the Hamiltonian and therefore they are stable in the Lyapunov sense, i.e. stable with respect to not only small perturbations but also against finite ones.  In particular, the boundedness of $H$ from below can  be viewed if one considers the scaling transformation $\psi=(1/a)^{1/2}\psi_{s}(x/a)$  retaining the number of waves $N$ where $\psi=\psi_{s}(x)$ is the soliton solution. Under this transform $H$ becomes a function of the scaling parameter $a$: 
\begin{equation}
H\left( a\right) =\left( \frac 1a-\frac 1{2a^2}\right) \frac{\mu}{2}\int |\psi_{s}|^{4}\,dx. \label{ham-scaling}
\end{equation}
It is worth noting that the dispersion term and all nonlinear terms in $H$, except $\int \frac{\mu}{2}|\psi|^{4}dx$, have the same scaling dependence $\propto a^{-2}$. The latter means that at $\mu=0$ Eq. (\ref{GNLSE}) can be related to the critical NLS equation like the two-dimensional cubic NLS equation.   
>From Eq. (\ref{ham-scaling}) it is also seen that for $\mu<0$ $H(a)$ has a minimum corresponding to the soliton.
Unlike the supercritical case, the scaling transformation for the other soliton  branch with $\mu>0$  gives a maximum of $H(a)$ on solitons and unboundedness of $H$ as $a\to 0$.  Under  the gauge transformation $\psi =\psi_{s}e^{i\chi }$,  on the contrary, 
the Hamiltonian reaches a minimum on soliton solutions and consequently the solitons  with $\mu>0$ represent saddle points. 
This indicates a possible instability of solitons for the whole subcritical branch, at least
with respect to finite perturbations. In this paper we investigate this question in more details. The main attention will be paid to the nonlinear stage of the instability. This problem, indeed, is not trivial in spite of a closed similarity with the critical NLSE. It is worth noting that Eq. (\ref{GNLSE}) at $\mu=\gamma=C=0$ represents an integrable model (the so-called derivative NLSE) \cite{KN}
and exponentially decaying solitons in this model  are stable. It is more or less evident also that  small coefficients  $\gamma,\,C$ cannot break the stability of solitons. This means that in the space of parameters we may expect the existence of a threshold. Above this threshold solitons must be unstable and the development of this instability may lead  to collapse, i.e. the formation of a singularity in finite time.
\\

{\bf 2}. 
Consider the Hamiltonian (\ref{hamiltonian})  written in terms of amplitude $r$ and phase $\varphi$ 
($\psi=re^{i\varphi}$):
\begin{equation} \label{ham}
H=\int \left[ r_{x}^2+\frac{\mu}{2}r^4-\frac{\gamma}{2}r^{2}\widehat{k}r^2-\frac 13r^6+r^2\left( \varphi _x+\beta r^2\right) ^2\right] dx,
\end{equation}
where by an appropriate choice of the new dimensionless variables the renormalized constant $\tilde C=C+\beta^2$ can be taken equal to 1.
Hence one can see that  the  Hamiltonian takes its minimum value when the last term in (\ref{ham}) vanishes, i.e.  when  
\begin{equation} \label{phase}
 \varphi _x+\beta r^2=0.
\end{equation} 
Integrating this equation gives an $x-$dependence for the phase, called chirp in nonlinear optics. It is interesting to note that the remaining part of the  Hamiltonian does not contain the phase at all. 
 
 First investigate the local model when $\gamma=0$.   Let the Hamiltonian be negative in some region $\Omega$ : $H_{\Omega}<0$. Then, following Refs. \cite{zakh, chaos}, one can establish that due to radiation of small amplitude waves $H_{\Omega}<0$ can only decrease, becoming more and more negative, but the maximum
 value of $|\psi|$, according to the mean value theorem, can only increase:
 \begin{equation} \label{rad-ineq}
 \max_{x\in \Omega}|\psi|^4\geq\frac{3|H_{\Omega}|}{N_{\Omega}}.
 \end{equation}
This process is possible only for Hamiltonians which are unbounded from below. In accordance with  (\ref{ham-scaling})
such a situation is realized when $\mu >0$. In this case the radiation leads to the appearance of infinitely large amplitudes $r$. However, it is impossible to conclude that the singularity formation develops in finite time.

For $\gamma>0$ the estimations on the maximum value of $|\psi|$ are not as transparent as they are for the local case. Instead of (\ref{rad-ineq}), it is possible to obtain a similar estimate,
$$
\max_x |\psi|^{4}\geq\frac{3|H|}{N}.
$$
However, it is expressed through the total Hamiltonian $H$ and the total number of waves $N$. Besides, two inequalities must be satisfied: $H<0$ and  $N<\frac{2N_{2}}{\gamma}$. For interfacial waves, $N_{2}\approx 1.39035> N_{cr}
\approx 1.3521$. Thus, the maximum amplitude in this case is bounded from below by a conservative quantity and this maximum can never disappear during the nonlinear evolution.     
   
Now we consider the situation  where the self-steepening process can be neglected ($\beta=0$). In this case Eq. (\ref{GNLSE}) 
becomes
\begin{eqnarray*}
i\psi_{t} + \psi_{xx} - \lambda^{2}\psi - \mu|\psi|^{2}\psi + \gamma\psi\widehat{k}|\psi|^{2}+3C|\psi|^{4}\psi =0.
\end{eqnarray*}
It is possible to obtain a criterion of collapse using the virial equation (for details, see \cite{VPT, zakh, Zakharov1984}). This equation is written for the positive definite quantity 
\begin{equation}
R = \int x^{2}|\psi|^{2}dx, \nonumber
\end{equation}
which, up to the multiplier $N$, coincides with the mean square size of the distribution. The second derivative of $R$   
with respect to time is defined by the virial equation
\begin{eqnarray}
R_{tt} = 8\left( H - \frac{\mu}{4}\int |\psi|^{4}dx \right). \label{virial}
\end{eqnarray}
Hence, for $\mu>0$ one can easily obtain the following inequality:
\begin{eqnarray}
R_{tt} < 8 H, \nonumber
\end{eqnarray}
which yields, after double integration, $R< 4Ht^{2}+\alpha_{1}t+\alpha_{2}$. Here $\alpha_{1,2}$ are constants which are
obtained from the initial conditions. Hence, it follows that for the states with negative Hamiltonian, $H<0$, there always exists such moment of time $t_0$ when the positive definite quantity $R$ vanishes. At this moment of time the amplitude becomes infinite. Therefore the condition $H<0$ represents a sufficient criterion of collapse (compare with \cite{VPT, zakh}).
However, it is necessary to add that this criterion can be improved by the same way as it was done in Refs. \cite{tur, KRRT} for the three-dimensional cubic NLS equation. From Eq. (\ref{virial}) one can see that for the stationary case (on the soliton solution) $H_s=
\frac{\mu}{4}\int |\psi_s|^{4}dx$, in agreement with Eq. (\ref{ham-scaling}). As we demonstrated before for $\mu>0$ the soliton realizes a saddle point of $H$ for fixed $N$. It follows from (\ref{ham-scaling}) that for small $a$ the Hamiltonian becomes unbounded from below, but for $a>1$ it decreases (this corresponds to spreading).  Therefore in order to achieve a blow-up regime the system should pass through the energetic barrier equal to $H_s$.  Thus, for this case the criterion $H<0$ must be changed into the sharper  criterion: $H<H_s$. This criterion can be obtained rigorously using step by step the scheme presented in \cite{tur, KRRT} and therefore we skip its derivation.
\\

{\bf 3.} 
In order to verify all the theoretical arguments about the formation of collapse presented above we performed a numerical integration of the NLSE (\ref{GNLSE}) for
$\mu >0$  by using the standard 4th order Runge-Kutta scheme. The initial conditions were chosen in the form of solitons but with larger amplitudes than for the stationary solitons.  The increase in initial amplitude was varied in the interval from 0.1\% up to 10\%.  The initial phase was given by means of Eq. (\ref{phase}). In all runs with theses initial conditions we observed a high increase of the soliton amplitude up to a factor 14 with a shrinking of its width. In the peak region  pulses for both IW and WW cases behaved similarly. Near the maximum the pulse peak was almost symmetric: anisotropy was not visible. The difference was observed in the asymptotic regions far from the pulse core where the pulses had different asymmetries for IW and WW because of the opposite sign for $\beta$ (see Eqs. (\ref{IW}), (\ref{WW})). For the given values of $\beta$ we did not observe the simultaneous formation of two types of singularities with blowing-up amplitudes and sharp gradients as it was demonstrated in the rcent numerical experiments for the three-dimensional collapse of short optical pulses due to self-focusing and self-steepening in the framework of the generalized NLS equation \cite{litvak} and equations of the Kadomtsev-Petvishvili type \cite{litvak1}. 

In our numerical computations we found that the amplitude and its spatial collapsing distribution develop in a self-similar manner.  Near the collapse point in the equation (with $\mu >0$) one can neglect  the term proportional to $\mu$. In this asymptotic regime Eq. (\ref{GNLSE}) admits self-similar solutions,
\begin{equation} 
\label{asympt}
r(x,t)=(t_0-t)^{-1/4}f\left(\frac{x}{(t_0-t)^{1/2}}\right),
\end{equation}
where $t_0$ is the collapse time.

\begin{figure}[tbp]
\begin{center}
\includegraphics[width=9cm,height=7cm]{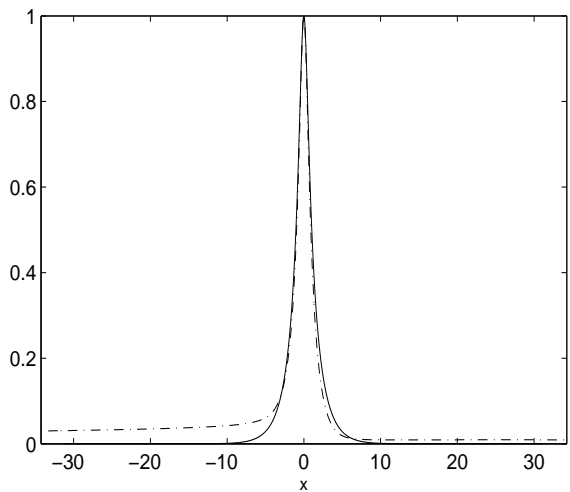}
\caption{Initial (solid line) and final (dashed line) at $t=1.18$ distributions for $|\overline{\psi}|$, interfacial waves, self-similar variables. The soliton amplitude was increased by 1\%, $\mu=1$, $\lambda=1$. The ratio between final and initial soliton amplitudes in the physical variables is about 11.}
\end{center}
\end{figure}

\begin{figure}[tbp]
\begin{center}
\includegraphics[width=9cm,height=7cm]{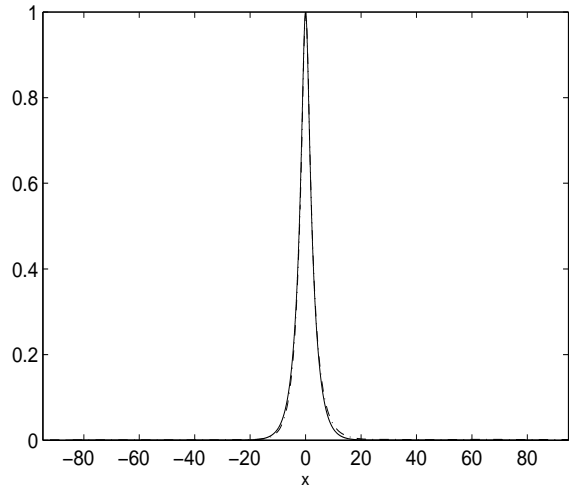}
\caption{Initial (solid line) and final (dashed line) at $t=2.7192$ distributions for $|\overline{\psi}|$, WW solitons, self-similar variables. The soliton amplitude was increased by 1\%, $\mu=1$, $\lambda=1$. The ratio between final and initial soliton amplitudes in the physical variables is about 11.}
\end{center}
\end{figure}

\begin{figure}[tbp]
\begin{center}
\includegraphics[width=9cm,height=7cm]{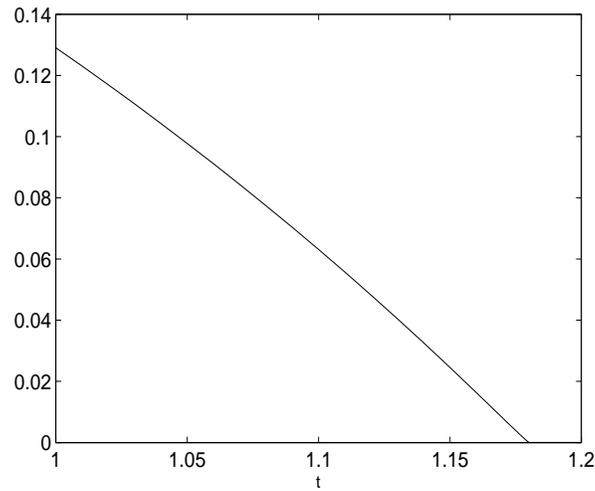}
\caption{Dependence of $1/\max|\psi|^{4}$ on time. Interfacial waves.}
\end{center}
\end{figure}

To verify that we approached the asymptotic behavior given by Eq. (\ref{asympt}),  we normalized at each moment of time the $\psi-$function
by  the maximum (in $x$) of its modulus $\max|\psi|\equiv M$ and introduced new self similar variables,
\begin{equation}
\psi(x,t)=M\overline{\psi}(\xi,\tau),\,\, \xi=M^2(x-x_{max}),\,\,\,\tau=\mbox{ln}\,M. \label{self-similarVar}
\end{equation}
Here $x_{max}$ is the point corresponding to the maximum of $|\psi|$. 
In comparison with those given by Eq. (\ref{asympt}), such new variables are more convenient because they do not require the determination of the collapsing time $t_0$. 

Fig. 1 and Fig. 2 show typical dependences of $|\overline{\psi}|$ as a function of the self-similar variable $\xi$ at $t=0$ (solid line) and at the final time (dashed line) for both the IW and WW cases. 
In both figures one can see a fairly good coincidence 
between the initial soliton distribution and the final one at the 
the central (collapsing) part of the pulse and asymmetry of the pulse at its tails due to self-steepening. The latter demonstrates that collapse has a self-similar behavior. The form of the central part of the pulse approaches the soliton shape because asymptotically the NLS model (\ref{GNLSE}) tends to the critical NLS system. It is necessary to mention that this has been well-known for the classical two-dimensional NLS equation since the paper by Fraiman \cite{fraiman}.

Fig. 3 shows how  $1/\max|\psi|^{4}$ depends on time. This dependence is almost linear in the correspondence with the self-similar law (\ref{asympt}). If the initial amplitudes were less than the stationary soliton values, then the distribution would spread in time dispersively, in full correspondence with qualitative arguments based on the scaling transformations (\ref{ham-scaling}). 
\\

{\bf 4.} The authors thank A.I. Dyachenko for valuable discussions
concerning numerical simulations. This paper was performed in the
framework of the NATO Linkage Grant EST.CLG.978941. The work of DA and EK
was also supported by RFBR (grants 06-01-00665, 07-01-92165), the Program of RAS ''Fundamental problems in nonlinear dynamics'' and Grant NSh 7550.2006.2.

\end{document}